\documentclass[a4paper,11pt]{article}

\makeatletter

\gdef\@fpheader{\newline}
\gdef\@journal{jhep}

\RequirePackage{amsmath}
\RequirePackage{amssymb}
\RequirePackage{epsfig}
\RequirePackage{CJK}
\RequirePackage{graphicx}
\RequirePackage[numbers,sort&compress]{natbib}
\RequirePackage{color}
\RequirePackage[dvipdfmx
,colorlinks=true
,urlcolor=blue
,anchorcolor=blue
,citecolor=blue
,filecolor=blue
,linkcolor=blue
,menucolor=blue
,pagecolor=blue
,linktocpage=true
,pdfproducer=medialab
,pdfa=true
,CJKbookmarks
]{hyperref}

\newif\ifnotoc\notocfalse
\newif\ifemailadd\emailaddfalse
\newif\iftoccontinuous\toccontinuousfalse

\def\@subheader{\@empty}
\def\@keywords{\@empty}
\def\@abstract{\@empty}
\def\@xtum{\@empty}
\def\@dedicated{\@empty}
\def\@arxivnumber{\@empty}
\def\@collaboration{\@empty}
\def\@collaborationImg{\@empty}
\def\@proceeding{\@empty}
\def\@preprint{\@empty}

\newcommand{\subheader}[1]{\gdef\@subheader{#1}}
\newcommand{\keywords}[1]{\if!\@keywords!\gdef\@keywords{#1}\else%
\PackageWarningNoLine{\jname}{Keywords already defined.\MessageBreak Ignoring last definition.}\fi}
\renewcommand{\abstract}[1]{\gdef\@abstract{#1}}
\newcommand{\dedicated}[1]{\gdef\@dedicated{#1}}
\newcommand{\arxivnumber}[1]{\gdef\@arxivnumber{#1}}
\newcommand{\proceeding}[1]{\gdef\@proceeding{#1}}
\newcommand{\xtumfont}[1]{\textsc{#1}}
\newcommand{\correctionref}[3]{\gdef\@xtum{\xtumfont{#1} \href{#2}{#3}}}
\newcommand\jname{JHEP}
\newcommand\acknowledgments{\section*{Acknowledgments}}

\newcommand\preprint[1]{\gdef\@preprint{\hfill #1}}

\newcommand\note[2][]{%
\if!#1!%
\stepcounter{footnote}\footnotetext{#2}%
\else%
{\renewcommand\thefootnote{#1}%
\footnotetext{#2}}%
\fi}

\newtoks\auth@toks
\renewcommand{\author}[2][]{%
  \if!#1!%
    \auth@toks=\expandafter{\the\auth@toks#2\ }%
  \else
    \auth@toks=\expandafter{\the\auth@toks#2$^{#1}$\ }%
  \fi
}

\newtoks\affil@toks\newif\ifaffil\affilfalse
\newcommand{\affiliation}[2][]{%
\affiltrue
  \if!#1!%
    \affil@toks=\expandafter{\the\affil@toks{\item[]#2}}%
  \else
    \affil@toks=\expandafter{\the\affil@toks{\item[$^{#1}$]#2}}%
  \fi
}

\newtoks\email@toks\newcounter{email@counter}%
\newcommand{\emailAdd}[1]{%
\emailaddtrue%
\ifnum\theemail@counter>0\email@toks=\expandafter{\the\email@toks, \@email{#1}}%
\else\email@toks=\expandafter{\the\email@toks\@email{#1}}%
\fi\stepcounter{email@counter}}
\newcommand{\@email}[1]{\href{mailto:#1}{\tt #1}}

\newcommand*\collaboration[1]{\gdef\@collaboration{#1}}
\newcommand*\collaborationImg[2][]{\gdef\@collaborationImg{#2}}

\newcommand\afterLogoSpace{\smallskip}
\newcommand\afterSubheaderSpace{\vskip3pt plus 2pt minus 1pt}
\newcommand\afterProceedingsSpace{\vskip21pt plus0.4fil minus15pt}
\newcommand\afterTitleSpace{\vskip23pt plus0.06fil minus13pt}
\newcommand\afterRuleSpace{\vskip23pt plus0.06fil minus13pt}
\newcommand\afterCollaborationSpace{\vskip3pt plus 2pt minus 1pt}
\newcommand\afterCollaborationImgSpace{\vskip3pt plus 2pt minus 1pt}
\newcommand\afterAuthorSpace{\vskip5pt plus4pt minus4pt}
\newcommand\afterAffiliationSpace{\vskip3pt plus3pt}
\newcommand\afterEmailSpace{\vskip16pt plus9pt minus10pt\filbreak}
\newcommand\afterXtumSpace{\par\bigskip}
\newcommand\afterAbstractSpace{\vskip16pt plus9pt minus13pt}
\newcommand\afterKeywordsSpace{\vskip16pt plus9pt minus13pt}
\newcommand\afterArxivSpace{\vskip3pt plus0.01fil minus10pt}
\newcommand\afterDedicatedSpace{\vskip0pt plus0.01fil}
\newcommand\afterTocSpace{\bigskip\medskip}
\newcommand\afterTocRuleSpace{\bigskip\bigskip}
\newlength{\affiliationsSep}
\newcommand\beforetochook{\pagestyle{myplain}\pagenumbering{roman}}

\DeclareFixedFont\trfont{OT1}{phv}{b}{sc}{11}

\renewcommand\maketitle{
\pagestyle{empty}
\thispagestyle{titlepage}
\setcounter{page}{0}
\noindent{\small\scshape\@fpheader}\@preprint\par
\afterLogoSpace
\if!\@subheader!\else\noindent{\trfont{\@subheader}}\fi
\afterSubheaderSpace
\if!\@proceeding!\else\noindent{\sc\@proceeding}\fi
\afterProceedingsSpace
{\LARGE\flushleft\sffamily\bfseries\@title\par}
\afterTitleSpace
\hrule height 1.5\p@%
\afterRuleSpace
\if!\@collaboration!\else
{\Large\bfseries\sffamily\raggedright\@collaboration}\par
\afterCollaborationSpace
\fi
\if!\@collaborationImg!\else
{\normalsize\bfseries\sffamily\raggedright\@collaborationImg}\par
\afterCollaborationImgSpace
\fi
{\bfseries\raggedright\sffamily\the\auth@toks\par}
\afterAuthorSpace
\ifaffil\begin{list}{}{%
\setlength{\leftmargin}{0.28cm}%
\setlength{\labelsep}{0pt}%
\setlength{\itemsep}{\affiliationsSep}%
\setlength{\topsep}{-\parskip}}
\itshape\small%
\the\affil@toks
\end{list}\fi
\afterAffiliationSpace
\ifemailadd 
\noindent\hspace{0.28cm}\begin{minipage}[l]{.9\textwidth}
\begin{flushleft}
\textit{E-mail:} \the\email@toks
\end{flushleft}
\end{minipage}
\else 
\PackageWarningNoLine{\jname}{E-mails are missing.\MessageBreak Plese use \protect\emailAdd\space macro to provide e-mails.}
\fi
\afterEmailSpace
\if!\@xtum!\else\noindent{\@xtum}\afterXtumSpace\fi
\if!\@abstract!\else\noindent{\renewcommand\baselinestretch{.9}\textsc{Abstract:}}\ \@abstract\afterAbstractSpace\fi
\if!\@keywords!\else\noindent{\textsc{Keywords:}} \@keywords\afterKeywordsSpace\fi
\if!\@arxivnumber!\else\noindent{\textsc{ArXiv ePrint:}} \href{http://arxiv.org/abs/\@arxivnumber}{\@arxivnumber}\afterArxivSpace\fi
\if!\@dedicated!\else\vbox{\small\it\raggedleft\@dedicated}\afterDedicatedSpace\fi
\ifnotoc\else
\iftoccontinuous\else\newpage\fi
\beforetochook\hrule
\tableofcontents
\afterTocSpace
\hrule
\afterTocRuleSpace
\fi
\setcounter{footnote}{0}
\pagestyle{myplain}\pagenumbering{arabic}
} 

\renewcommand{\baselinestretch}{1.1}\normalsize
\divide\@tempdima\baselineskip
\@tempcnta=\@tempdima
\skip\@mpfootins = \skip\footins
\renewcommand{\@dotsep}{10000}

\newcommand\ps@myplain{
\pagenumbering{arabic}
\renewcommand\@oddfoot{\hfill-- \thepage\ --\hfill}
\renewcommand\@oddhead{}}
\let\ps@plain=\ps@myplain

\newcommand\ps@titlepage{\renewcommand\@oddfoot{}\renewcommand\@oddhead{}}

\numberwithin{equation}{section}

\renewcommand\section{\@startsection{section}{1}{\z@}%
                                   {-3.5ex \@plus -1.3ex \@minus -.7ex}%
                                   {2.3ex \@plus.4ex \@minus .4ex}%
                                   {\normalfont\large\bfseries}}
\renewcommand\subsection{\@startsection{subsection}{2}{\z@}%
                                   {-2.3ex\@plus -1ex \@minus -.5ex}%
                                   {1.2ex \@plus .3ex \@minus .3ex}%
                                   {\normalfont\normalsize\bfseries}}
\renewcommand\subsubsection{\@startsection{subsubsection}{3}{\z@}%
                                   {-2.3ex\@plus -1ex \@minus -.5ex}%
                                   {1ex \@plus .2ex \@minus .2ex}%
                                   {\normalfont\normalsize\bfseries}}
\renewcommand\paragraph{\@startsection{paragraph}{4}{\z@}%
                                   {1.75ex \@plus1ex \@minus.2ex}%
                                   {-1em}%
                                   {\normalfont\normalsize\bfseries}}
\renewcommand\subparagraph{\@startsection{subparagraph}{5}{\parindent}%
                                   {1.75ex \@plus1ex \@minus .2ex}%
                                   {-1em}%
                                   {\normalfont\normalsize\bfseries}}

\def\fnum@figure{\textbf{\figurename\nobreakspace\thefigure}}
\def\fnum@table{\textbf{\tablename\nobreakspace\thetable}}

\long\def\@makecaption#1#2{%
  \vskip\abovecaptionskip
  \sbox\@tempboxa{\small #1. #2}%
  \ifdim \wd\@tempboxa >\hsize
    \small #1. #2\par
  \else
    \global \@minipagefalse
    \hb@xt@\hsize{\hfil\box\@tempboxa\hfil}%
  \fi
  \vskip\belowcaptionskip}

\renewenvironment{thebibliography}[1]{%
\begin{oldthebibliography}{#1}%
\small%
\raggedright%
\setlength{\itemsep}{5pt plus 0.2ex minus 0.05ex}%
}%
{%
\end{oldthebibliography}%
}

\makeatother

\usepackage[T1]{fontenc} 
\usepackage{romannum} 
\usepackage{CJK}

\begin{document} 
\begin{CJK*}{GBK}{song}


\title{{\boldmath Geodesic dual spacetime} \\  }

\author[a,b]{Wen-Du Li}
\author[b*]{and Wu-Sheng Dai}\note{daiwusheng@tju.edu.cn.}


\affiliation[a]{Theoretical Physics Division, Chern Institute of Mathematics, Nankai University, Tianjin, 300071, P. R. China}
\affiliation[b]{Department of Physics, Tianjin University, Tianjin 300350, P.R. China}













\abstract{A duality between spacetime manifolds, the geodesic duality, is introduced.
Two manifolds are geodesic dual, if the transformation between their metrics
is also the transformation between their geodesics. That is, the
transformation that transforms the metric to the metric of the dual manifold
is also the transformation that transforms the geodesic to the geodesic of the
dual manifold. On the contrary, for nondual spacetime manifolds, a geodesic is
no longer a geodesic after the transformation between the metrics. We give a
general result of the duality between spacetime manifolds with diagonal
metrics. The geodesic duality of spherically symmetric spacetime are discussed
for illustrating the concept. The geodesic dual spacetime of the Schwarzschild
spacetime and the geodesic dual spacetime of the Reissner-Nordstr\"{o}m
spacetime are presented.}




\maketitle 

\flushbottom

\section{Introduction}

Geodesics, the trajectory of a free particle, reflects the geometry of
spacetime. Generally speaking, a transformation between the metrics of two
spacetimes cannot transform the geodesic of one spacetime to the geodesic of
another spacetime. In this paper, we show that there exist spacetimes whose
geodesics can be transformed to each other by the transformation between the
metrics of spacetimes. This is a duality between spacetimes, the geodesic duality.

By spacetime with diagonal metrics, we illustrate the concept of the geodesic
duality. The spherically symmetric spacetime is generally discussed. The dual
spacetime of the Schwarzschild spacetime and dual spacetime of the
Reissner-Nordstr\"{o}m spacetime are presented.

Duality bridges superficially different things. An important duality is the
AdS/CFT
duality\ \cite{maldacena1997large,witten1998anti,witten1998anti2,aharony2000large,d2004supersymmetric}%
. There exists dualities between spacetime manifolds and fluids, i.e., the
gravity/Fluid dual \cite{bredberg2012navier,hubeny2011fluid}, such as the
duality between $d+2$-dimensional Ricci-flat metrics and $d+1$-dimensional
fluids \cite{compere2011holographic,hao2015flat}, the geometrical duality of
both Brans-Dicke theory and general relativity \cite{quiros2000dual}, the
duality between the nonlinear equations of boundary fluid and gravity theory
\cite{bhattacharyya2008nonlinear,bhattacharyya2009forced}, the duality between
the conformal Navier Stokes equation and the long wavelength solution of
gravity
\cite{bhattacharyya2008conformal,ashok2014forced,bhattacharyya2009incompressible}%
, the duality between $d+2$-dimensional Ricci-flat metrics and $d+1$%
-dimensional relativistic fluids \cite{compere2012relativistic}, a dual
relation of $d+2$-dimensional metrics corresponding to $d+1$-dimensional
fluids \cite{pinzani2015towards}, and the fluid/gravity dual in spacetime with
general non-rotating weakly isolated horizons \cite{wu2013fluid}. The
gravoelectric duality decomposes the Riemann curvature into electric and
magnetic parts and introduces the gravoelectric duality transformation by
interchange of active and passive electric parts which amounts to interchange
of the Ricci and Einstein tensors. The gravoelectric duality can be used to
find the solution of the Einstein equation
\cite{dadhich2002most,nouri1999spacetime,dadhich2000electromagnetic,dadhich1998duality}%
. Invariances under the gravoelectric duality are also considered
\cite{dadhich2000gravoelectric}. There are various dualities between physical
systems, such as the duality between the two-dimensional Ising and planar
Heisenberg models to gauge theories in four dimensions \cite{sugamoto1979dual}%
, the duality between the $SU(2)$ Higgs-Kibble model and the relativistic
hydrodynamics of Freedman coupled to Higgs scalars \cite{seo1979dual}, the
quantum equivalence of dual field theories \cite{fradkin1985quantum}, the
duality in many-component Ising models in two dimensions on a square lattice
\cite{mittag1971dual}.

In section \ref{generaldiscussion}, we define the concept of the geodesic
duality. In section \ref{spherically}, we give a general discussion on the
spherically symmetric geodesic dual spacetime. In sections \ref{Schwarzschild}
and \ref{R-N}, we give the dual spacetime of the Schwarzschild spacetime and
the Reissner-Nordstr\"{o}m spacetime.

\section{Geodesic duality \label{generaldiscussion}}

The geodesic duality is defined as follows.

\textit{Two manifolds are geodesic dual, if the transformation between their
metrics is also the transformation between their geodesics. }

We illustrate the concept of the geodesic duality through examples of
spacetime manifolds with diagonal metrics.

A geodesic on an $n+1$-dimensional manifold is determined by $n+1$ geodesic
equations. A geodesic on a manifold, after a transformation between the
metrics of two manifolds, will be transformed to a curve on the other
manifold. If these two manifolds are geodesic dual, the curve will still be a geodesic.

In the following, we consider the geodesic duality between two spacetime
manifolds with diagonal metrics. The idea apples also to more general cases.

$n+1$ geodesic equations, which determine an $n+1$-dimensional geodesic, have
$n+1$ first integrals, say $Q_{0}$, $Q_{1}$,$\ldots$, $Q_{n}$. Of its $n+1$
first integrals, suppose that $n$ first integrals, $Q_{0}$, $\cdots$,
$Q_{\eta-1}$, $Q_{\eta+1}$,$\ldots$, $Q_{n}$, are known and one first integral
$Q_{\eta}$ is left unknown. For a diagonal metric%
\begin{equation}
ds^{2}=g_{\mu\mu}\left(  dx^{\mu}\right)  ^{2}, \label{metric1}%
\end{equation}
the geodesic is given by
\begin{equation}
\frac{dx^{\lambda}}{dx^{\eta}}=\frac{1/g_{\lambda\lambda}}{\sqrt{\frac
{1}{g_{\eta\eta}Q_{\lambda}^{2}}-\frac{1}{g_{\eta\eta}}\left(  \frac{Q_{0}%
^{2}}{Q_{\lambda}^{2}}\frac{1}{g_{00}}+\cdots+\frac{Q_{a}^{2}}{Q_{\lambda}%
^{2}}\frac{1}{g_{aa}}+\ldots+\frac{Q_{\eta-1}^{2}}{Q_{\lambda}^{2}}\frac
{1}{g_{\eta-1\eta-1}}+\frac{Q_{\eta+1}^{2}}{Q_{\lambda}^{2}}\frac{1}%
{g_{\eta+1\eta+1}}+\ldots+\frac{1}{g_{\lambda\lambda}}+\ldots+\frac{Q_{n}^{2}%
}{Q_{\lambda}^{2}}\frac{1}{g_{nn}}\right)  }}, \label{geodesic}%
\end{equation}
where $\lambda\neq\eta$.

Geodesic duality means that after a duality transformation, Eq.
(\ref{geodesic}), the geodesic equation of the manifold with the metric
(\ref{metric1}), must also be a geodesic equation of its geodesic dual
manifold. The geodesic equation on the dual manifold is still of the form of
Eq. (\ref{geodesic}). To obtain the geodesic equation of the dual manifold, we
exchange the term $\frac{1}{g_{\lambda\lambda}}$ with another term in the
square root of Eq. (\ref{geodesic}):
\begin{equation}
\frac{Q_{a}^{2}}{Q_{\lambda}^{2}}\frac{1}{g_{aa}}\Leftrightarrow\frac
{1}{g_{\lambda\lambda}}. \label{exchange}%
\end{equation}
We then arrive at a new equation. Requiring the new equation to be a geodesic
equation described by the metric $\widetilde{g}_{\mu\nu}$ gives a
transformation%
\begin{align}
x^{\mu}  &  =\chi^{\mu}\left(  y^{\nu}\right)  ,\label{x}\\
Q_{\lambda}  &  =\Omega_{\lambda}\left(  q_{\kappa}\right)  \label{Q}%
\end{align}
determined by%
\begin{equation}
\frac{\Omega_{a}^{2}\left(  q_{\kappa}\right)  }{\Omega_{\lambda}^{2}\left(
q_{\kappa}\right)  }\frac{1}{g_{aa}\left(  \chi^{\mu}\left(  y^{\nu}\right)
\right)  }=\frac{1}{g_{\lambda\lambda}\left(  y^{\nu}\right)  }, \label{trans}%
\end{equation}
where\ $y^{\nu}$ is the coordinate of the manifold described by $\widetilde{g}%
_{\mu\nu}$, and $q_{\kappa}$ is the first integral of the new geodesic
equation. Rewriting the new geodesic equation in the form of the geodesic
equation (\ref{geodesic}):%
\begin{equation}
\frac{dy^{\lambda}}{dy^{\eta}}=\frac{1/\tilde{g}_{\lambda\lambda}}{\sqrt
{\frac{1}{\tilde{g}_{\eta\eta}q_{\lambda}^{2}}-\frac{1}{\tilde{g}_{\eta\eta}%
}\left(  \frac{q_{0}^{2}}{q_{\lambda}^{2}}\frac{1}{\tilde{g}_{00}}%
+\cdots+\frac{1}{\tilde{g}_{aa}}+\ldots+\frac{q_{\eta-1}^{2}}{q_{\lambda}^{2}%
}\frac{1}{\tilde{g}_{\eta-1\eta-1}}+\frac{q_{\eta+1}^{2}}{q_{\lambda}^{2}%
}\frac{1}{\tilde{g}_{\eta+1\eta+1}}+\ldots+\frac{q_{\lambda}^{2}}{q_{\lambda
}^{2}}\frac{1}{\tilde{g}_{\lambda\lambda}}+\ldots+\frac{q_{n}^{2}}{q_{\lambda
}^{2}}\frac{1}{\tilde{g}_{nn}}\right)  }}, \label{newgoedesic}%
\end{equation}
where $\lambda\neq\eta$, we can read out the metric of the dual manifold,
$\widetilde{g}_{\mu\nu}$, directly. It can be found from Eq.
(\ref{newgoedesic}) that the metric of the dual manifold, $\widetilde{g}%
_{\mu\nu}$, is also diagonal, i.e.,
\begin{equation}
d\sigma^{2}=\tilde{g}_{\mu\mu}\left(  dy^{\mu}\right)  ^{2}. \label{metric2}%
\end{equation}
The manifold with the metric $\tilde{g}_{\mu\nu}$ is the geodesic duality of
the manifold with the metric $g_{\mu\nu}$.

In the following, we will illustrate how to solve the duality transformations
(\ref{x}) and (\ref{Q}) through examples.

It should be emphasized that an $n+1$-dimensional spacetime has $n-1$ geodesic
dual spacetimes. This is because the number of the term\ $\frac{Q_{a}^{2}%
}{Q_{\lambda}^{2}}\frac{1}{g_{aa}}$ is $n-1$, and then there are $n-1$ ways to
exchange the term $\frac{1}{g_{\lambda\lambda}}$ with the terms $\frac
{Q_{a}^{2}}{Q_{\lambda}^{2}}\frac{1}{g_{aa}}$.

\section{Geodesic duality: spherically symmetric spacetime \label{spherically}%
}

In this section, we consider spherically symmetric spacetime.

\subsection{Duality transformation}

The metric of a spherically symmetric spacetime is of the form
\begin{equation}
ds^{2}=g_{00}\left(  x^{1}\right)  \left(  dx^{0}\right)  ^{2}+g_{11}\left(
x^{1}\right)  \left(  dx^{1}\right)  ^{2}+\left(  x^{1}\right)  ^{2}\left[
\left(  dx^{2}\right)  ^{2}+\sin^{2}x^{2}\left(  dx^{3}\right)  ^{2}\right]
.\label{ds1}%
\end{equation}
In the spherical coordinate $x^{0}=t$, $x^{1}=r$, $x^{2}=\theta$,
$x^{3}=\varphi$ and then $g_{00}=-f\left(  r\right)  $, $g_{11}=g\left(
r\right)  $, $g_{22}=r^{2}$, and $g_{33}=r^{2}\sin^{2}\theta$ with $f\left(
r\right)  $ and $g\left(  r\right)  $ the functions of $r$.

The geodesic equation on spherically symmetric spacetime, by Eq.
(\ref{geodesic}), is
\begin{equation}
\frac{dx^{3}}{dx^{1}}=\frac{1}{g_{33}\sqrt{\frac{1}{g_{11}Q_{3}^{2}}-\frac
{1}{g_{11}}\left(  \frac{Q_{0}^{2}}{Q_{3}^{2}}\frac{1}{g_{00}}+\frac{1}%
{g_{33}}\right)  }}.\label{geo11}%
\end{equation}
In spherically symmetric cases the geodesic orbit is in a plane, so we can
choose $x^{2}=\pi/2$, then, $g_{33}=\left(  x^{1}\right)  ^{2}$ and the first
integral $Q_{2}=0$ which is the $x^{2}$-component of the angular momentum.

In this case, the only possible exchange in Eq. (\ref{exchange}) is
\begin{equation}
\frac{Q_{0}^{2}}{Q_{3}^{2}}\frac{1}{g_{00}}\Leftrightarrow\frac{1}{g_{33}}.
\end{equation}
In spherically symmetric cases, only the radial coordinate $x^{1}=r$ needs to
be considered, so there is only one coordinate transformation
\begin{equation}
x^{1}=\chi^{1}\left(  y^{1}\right)  .
\end{equation}
The duality relation is determined by%
\begin{equation}
\frac{Q_{0}^{2}}{Q_{3}^{2}}\frac{1}{g_{00}\left(  \chi\left(  y^{1}\right)
\right)  }=\frac{1}{\left(  y^{1}\right)  ^{2}},
\end{equation}
so the duality relation (\ref{trans}) is%
\begin{equation}
x^{1}\rightarrow\chi^{1}\left(  y^{1}\right)  =g_{00}^{-1}\left(  \frac
{Q_{0}^{2}}{Q_{3}^{2}}\left(  y^{1}\right)  ^{2}\right)  .\label{rchirho1}%
\end{equation}

Note that the duality relation of the angular coordinate is trivial:
$x^{3}\rightarrow\beta y^{3}$.

Now consider the transformation of the angular coordinate. Suppose the
transformation is%
\begin{equation}
x^{3}\rightarrow\chi^{3}\left(  y^{1}\right)  y^{3}\left(  y^{1}\right)  .
\label{x3y3trans}%
\end{equation}
Substituting Eq. (\ref{x3y3trans}) into the geodesic equation (\ref{geo11})
gives%
\begin{equation}
\frac{dy^{3}\left(  y^{1}\right)  }{dx^{1}}=\frac{1}{\chi_{3}\left(
y^{1}\right)  g_{33}\sqrt{\frac{1}{g_{11}Q_{3}^{2}}-\frac{1}{g_{11}}\left(
\frac{Q_{0}^{2}}{Q_{3}^{2}}\frac{1}{g_{00}}+\frac{1}{g_{33}}\right)  }}%
-\frac{y^{3}\left(  y^{1}\right)  }{\chi_{3}\left(  y^{1}\right)  }\frac
{d\chi_{3}\left(  y^{1}\right)  }{dx^{1}}. \label{Eqtrans}%
\end{equation}
If Eq. (\ref{Eqtrans}) is still a geodesic equation, we must have%
\begin{equation}
\frac{y^{3}\left(  y^{1}\right)  }{\chi_{3}\left(  y^{1}\right)  }\frac
{d\chi_{3}\left(  y^{1}\right)  }{dx^{1}}=0,
\end{equation}
so%
\begin{equation}
\chi_{3}\left(  y^{1}\right)  =\beta\label{chibeta}%
\end{equation}
with $\beta$ a constant.

The geodesic equation on the dual spacetime then can be obtained by submitting
the duality relation (\ref{rchirho1}) into the geodesic equation (\ref{geo11}):%

\begin{align}
\frac{dy^{3}}{dy^{1}}  &  =\frac{1}{\left(  y^{1}\right)  ^{2}}\left\{
\frac{1}{g_{11}\left(  g_{00}^{-1}\left(  \frac{Q_{0}^{2}}{Q_{3}^{2}}\left(
y^{1}\right)  ^{2}\right)  \right)  }\left\{  \frac{\beta\left[  g_{00}%
^{-1}\left(  \frac{Q_{0}^{2}}{Q_{3}^{2}}\left(  y^{1}\right)  ^{2}\right)
\right]  ^{2}}{\left(  y^{1}\right)  ^{2}\left[  g_{00}^{-1}\left(
\frac{Q_{0}^{2}}{Q_{3}^{2}}\left(  y^{1}\right)  ^{2}\right)  \right]
^{\prime}}\right\}  ^{2}\frac{1}{Q_{3}^{2}}\right. \nonumber\\
&  \left.  -\frac{1}{g_{11}\left(  g_{00}^{-1}\left(  \frac{Q_{0}^{2}}%
{Q_{3}^{2}}\left(  y^{1}\right)  ^{2}\right)  \right)  }\left\{  \frac
{\beta\left[  g_{00}^{-1}\left(  \frac{Q_{0}^{2}}{Q_{3}^{2}}\left(
y^{1}\right)  ^{2}\right)  \right]  ^{2}}{\left(  y^{1}\right)  ^{2}\left[
g_{00}^{-1}\left(  \frac{Q_{0}^{2}}{Q_{3}^{2}}\left(  y^{1}\right)
^{2}\right)  \right]  ^{\prime}}\right\}  ^{2}\left\{  \frac{1}{\left[  \alpha
g_{00}^{-1}\left(  \frac{Q_{0}^{2}}{Q_{3}^{2}}\left(  y^{1}\right)
^{2}\right)  \right]  ^{2}}\alpha^{2}+\frac{1}{\left(  y^{1}\right)  ^{2}%
}\right\}  \right\}  ^{-1/2}. \label{trans11}%
\end{align}

Rewriting the new equation to the standard form of the geodesic equation given
by Eq. (\ref{geodesic}),%
\begin{equation}
\frac{dy^{3}}{dy^{1}}=\frac{1}{\widetilde{g}_{33}\sqrt{\frac{1}{\widetilde{g}%
_{11}q_{3}^{2}}-\frac{1}{\widetilde{g}_{11}}\left(  \frac{q_{0}^{2}}{q_{3}%
^{2}\widetilde{g}_{00}}+\frac{1}{\widetilde{g}_{33}}\right)  }}%
,\label{geodesic21}%
\end{equation}
we can read out the metric of the dual spacetime:
\begin{align}
\widetilde{g}_{00}\left(  y^{1}\right)   &  =\left[  \alpha g_{00}^{-1}\left(
\frac{Q_{0}^{2}}{Q_{3}^{2}}\left(  y^{1}\right)  ^{2}\right)  \right]
^{2},\label{vqrg00}\\
\widetilde{g}_{11}\left(  y^{1}\right)   &  =g_{11}\left(  g_{00}^{-1}\left(
\frac{Q_{0}^{2}}{Q_{3}^{2}}\left(  y^{1}\right)  ^{2}\right)  \right)
\left\{  \frac{\left(  y^{1}\right)  ^{2}\left[  g_{00}^{-1}\left(
\frac{Q_{0}^{2}}{Q_{3}^{2}}\left(  y^{1}\right)  ^{2}\right)  \right]
^{\prime}}{\beta\left[  g_{00}^{-1}\left(  \frac{Q_{0}^{2}}{Q_{3}^{2}}\left(
y^{1}\right)  ^{2}\right)  \right]  ^{2}}\right\}  ^{2},\label{vqrg11}\\
q_{3} &  =Q_{3},\\
\frac{q_{0}}{q_{3}} &  =\alpha,
\end{align}
where we choose $y^{2}=\pi/2$, then $\widetilde{g}_{33}=\left(  y^{1}\right)
^{2}$, the first integral $q_{2}=0$ which is the $y^{2}$-component of the
angular momentum, and $\alpha$ is a constant. Then the metric of the dual
manifold is%
\begin{equation}
d\sigma^{2}=\widetilde{g}_{00}\left(  y^{1}\right)  \left(  dy^{0}\right)
^{2}+\widetilde{g}_{11}\left(  y^{1}\right)  \left(  dy^{1}\right)
^{2}+\left(  y^{1}\right)  ^{2}\left[  \left(  dy^{2}\right)  ^{2}+\sin
^{2}y^{2}\left(  dy^{3}\right)  ^{2}\right]  \label{dsigma1}%
\end{equation}
which is also spherically symmetric.

We might as well choose the constant%
\begin{equation}
\alpha=\frac{Q_{0}}{Q_{3}}%
\end{equation}
so that
\begin{align}
q_{0} &  =Q_{0},\\
q_{3} &  =Q_{3}.
\end{align}
Consequently, the duality relation between two geodesics on dual spacetimes
can be summarized as%
\begin{align}
Q_{3}\frac{g_{00}\left(  x^{1}\right)  }{Q_{0}} &  =q_{0}\frac{\tilde{g}%
_{33}\left(  y^{1}\right)  }{q_{3}},\label{dt11}\\
y^{3} &  =\beta x^{3}.\label{dt21}%
\end{align}

The geodesic on the spacetime (\ref{dsigma1}), Eq. (\ref{geodesic21}), can be
achieved directly from the geodesic on its geodesic dual spacetime
(\ref{ds1}), Eq. (\ref{geo11}), through the duality transformation
(\ref{dt11})-(\ref{dt21}). This result can be verified directly by
substituting the duality transforms (\ref{dt11})-(\ref{dt21}) into the
geodesic equation.

\subsection{Inverse duality transformation}

In addition, the inverse transformations of the duality transformations
(\ref{dt11}) and (\ref{dt21}) are
\begin{align}
y^{1}  &  \rightarrow\widetilde{g}_{00}^{-1}\left[  \frac{q_{0}^{2}}{q_{3}%
^{2}}\left(  x^{1}\right)  ^{2}\right]  ,\label{x1y1}\\
y^{3}  &  \rightarrow\frac{1}{\beta}x^{3}. \label{x3y3}%
\end{align}
By the transformations (\ref{x1y1}) and (\ref{x3y3}), Eqs. (\ref{vqrg00}) and
(\ref{vqrg11}) become%
\begin{align}
g_{00}\left(  x^{1}\right)   &  =\left[  \alpha\widetilde{g}_{00}^{-1}\left(
\frac{q_{0}^{2}}{q_{3}^{2}}\left(  x^{1}\right)  ^{2}\right)  \right]  ^{2},\\
g_{11}\left(  x^{1}\right)   &  =\widetilde{g}_{11}\left(  \widetilde{g}%
_{00}^{-1}\left(  \frac{q_{0}^{2}}{q_{3}^{2}}\left(  x^{1}\right)
^{2}\right)  \right)  \left\{  \frac{\left(  x^{1}\right)  ^{2}\beta\left[
\widetilde{g}_{00}^{-1}\left(  \frac{q_{0}^{2}}{q_{3}^{2}}\left(
x^{1}\right)  ^{2}\right)  \right]  ^{\prime}}{\left[  \widetilde{g}_{00}%
^{-1}\left(  \frac{q_{0}^{2}}{q_{3}^{2}}\left(  x^{1}\right)  ^{2}\right)
\right]  ^{2}}\right\}  ^{2},\\
Q_{3}  &  =q_{3},\\
\frac{Q_{0}}{Q_{3}}  &  =\alpha.
\end{align}
Then the geodesic equation (\ref{geodesic21}) returns to the geodesic equation
(\ref{geo11}).

\subsection{Duality}

To sum up, the duality relation between two geodesic dual spacetimes is%

\begin{align}
Q_{3}\frac{g_{00}\left(  x^{1}\right)  }{Q_{0}} &  =q_{0}\frac{\widetilde{g}%
_{33}\left(  y^{1}\right)  }{q_{3}},\\
q_{3}\frac{\widetilde{g}_{00}\left(  y^{1}\right)  }{q_{0}} &  =Q_{0}%
\frac{g_{33}\left(  x^{1}\right)  }{Q_{3}},\\
\frac{g_{11}\left(  x^{1}\right)  }{\left(  x^{1}\right)  ^{4}\left[
\widetilde{g}_{00}^{-1}\left(  \frac{q_{0}^{2}}{q_{3}^{2}}\left(
x^{1}\right)  ^{2}\right)  \right]  ^{\prime}} &  =\beta^{2}\frac
{\widetilde{g}_{11}\left(  y^{1}\right)  }{\left(  y^{1}\right)  ^{4}\left[
g_{00}^{-1}\left(  \frac{Q_{0}^{2}}{Q_{3}^{2}}\left(  y^{1}\right)
^{2}\right)  \right]  ^{\prime}}.
\end{align}
This duality transformation transforms both the metric and the geodesic to
that of the dual spacetime.

\subsection{Spherical coordinate form}

As a complement, we rewrite the above result with explicit spherical coordinates.

The metric (\ref{ds1}) and the geodesic equation (\ref{geo11}) under the
spherical coordinate, $x^{0}=t$, $x^{1}=r$, $x^{2}=\theta$, $x^{3}=\varphi$,
are%
\begin{equation}
ds^{2}=-f\left(  r\right)  dt^{2}+g\left(  r\right)  dr^{2}+r^{2}\left(
d\theta^{2}+\sin^{2}\theta d\varphi^{2}\right)  \label{ds}%
\end{equation}
and%
\begin{equation}
\frac{d\varphi}{dr}=\frac{1}{r^{2}}\frac{1}{\sqrt{\frac{1}{b^{2}}\frac
{1}{f\left(  r\right)  g\left(  r\right)  }-\frac{1}{g\left(  r\right)
}\left(  \frac{1}{a^{2}}+\frac{1}{r^{2}}\right)  }}, \label{geodesic1}%
\end{equation}
where $b^{2}=-Q_{3}^{2}/Q_{0}^{2}=L^{2}/E^{2}$ and $a^{2}=-Q_{3}^{2}%
=L^{2}/m^{2}$ with $L$ the angular momentum and $E$ the energy.

For the dual spacetime, the metric (\ref{dsigma1}) and the geodesic equation
(\ref{geodesic21}) under the spherical coordinate $y^{0}=\tau$, $y^{1}=\rho$,
$y^{2}=\vartheta$, $y^{3}=\phi$, become
\begin{equation}
d\sigma^{2}=-F\left(  \rho\right)  d\tau^{2}+G\left(  \rho\right)  d\rho
^{2}+\rho^{2}\left(  d\vartheta^{2}+\sin^{2}\vartheta d\phi^{2}\right)
\label{dsigma}%
\end{equation}
and%
\begin{equation}
\frac{d\phi}{d\rho}=\frac{1}{\rho^{2}}\frac{1}{\sqrt{\frac{1}{B^{2}}\frac
{1}{F\left(  \rho\right)  G\left(  \rho\right)  }-\frac{1}{G\left(
\rho\right)  }\left(  \frac{1}{A^{2}}+\frac{1}{\rho^{2}}\right)  }%
},\label{geodesic2}%
\end{equation}
where $B^{2}=-q_{3}^{2}/q_{0}^{2}=\ell^{2}/\mathcal{E}^{2}$ and $A^{2}%
=-q_{3}^{2}=\ell^{2}/m^{2}$ with $\ell$ the angular momentum and $\mathcal{E}$
the energy.

The duality relation here is%
\begin{align}
bf\left(  r\right)   &  =-\frac{\rho^{2}}{B},\label{dual1}\\
-\frac{r^{2}}{b}  &  =BF\left(  \rho\right)  ,\label{dual2}\\
\frac{g\left(  r\right)  }{r^{4}\left[  F^{-1}\left(  -\frac{r^{2}}{B^{2}%
}\right)  \right]  ^{\prime}}  &  =\beta^{2}\frac{G\left(  \rho\right)  }%
{\rho^{4}\left[  f^{-1}\left(  -\frac{\rho^{2}}{b^{2}}\right)  \right]
^{\prime}}. \label{dual3}%
\end{align}

\section{Geodesic dual spacetime of Schwarzschild spacetime
\label{Schwarzschild}}

For the Schwarzschild spacetime%
\begin{equation}
ds^{2}=-\left(  1-\frac{2M}{r}\right)  dt^{2}+\frac{1}{1-\frac{2M}{r}}%
dr^{2}+r^{2}\left(  d\theta^{2}+r^{2}\sin^{2}\theta d\varphi^{2}\right)  ,
\label{schw}%
\end{equation}
the duality transformations (\ref{rchirho1}) and (\ref{chibeta}) becomes
\begin{align}
r  &  \rightarrow\frac{2b^{2}M}{b^{2}+\rho^{2}},\\
\varphi &  \rightarrow\beta\phi.
\end{align}
Then the metric of the geodesic dual spacetime of the Schwarzschild spacetime
is
\begin{equation}
d\sigma^{2}=-\left(  \frac{2bM}{b^{2}+\rho^{2}}\right)  ^{2}d\tau^{2}+\left(
\frac{\rho^{2}}{\beta bM}\right)  ^{2}d\rho^{2}-\rho^{2}\left(  d\vartheta
^{2}+\sin^{2}\vartheta d\phi^{2}\right)  . \label{schwdual}%
\end{equation}
The geodesic equations is%
\begin{align}
\frac{d\varphi}{dr}  &  =\frac{1}{r^{2}}\frac{1}{\sqrt{\frac{1}{b^{2}}-\left(
1-\frac{2M}{r}\right)  \left(  \frac{1}{a^{2}}+\frac{1}{r^{2}}\right)  }},\\
\frac{d\phi}{d\rho}  &  =\frac{1}{\rho^{2}}\frac{1}{\sqrt{\frac{1}{B^{2}}%
\frac{1}{\left(  \frac{2bM}{b^{2}+\rho^{2}}\right)  ^{2}\left(  \frac{\rho
^{2}}{\beta bM}\right)  ^{2}}-\frac{1}{\left(  \frac{\rho^{2}}{\beta
bM}\right)  ^{2}}\left(  \frac{1}{A^{2}}+\frac{1}{\rho^{2}}\right)  }}.
\end{align}
It can be checked that Eqs. (\ref{schw}) and (\ref{schwdual}) satisfy the
duality relation (\ref{dual1})-(\ref{dual3}).

\section{Geodesic dual spacetime of Reissner-Nordstr\"{o}m spacetime
\label{R-N}}

For the Reissner-Nordstr\"{o}m spacetime%
\begin{equation}
d\tau^{2}=-\left(  1-\frac{2M}{r}+\frac{Q^{2}}{r^{2}}\right)  dt^{2}+\frac
{1}{1-\frac{2M}{r}+\frac{Q^{2}}{r^{2}}}dr^{2}+r^{2}d\theta^{2}+r^{2}\sin
^{2}\theta d\varphi^{2}, \label{RN}%
\end{equation}
the duality transformation (\ref{rchirho1}) becomes%
\begin{align}
r  &  \rightarrow\frac{b^{2}M+b^{2}\sqrt{\left(  M^{2}-Q^{2}\right)
-Q^{2}\frac{\rho^{2}}{b^{2}}}}{b^{2}+\rho^{2}},\\
\varphi &  \rightarrow\beta\phi.
\end{align}
The metric of the geodesic dual spacetime is then%
\begin{equation}
d\tau^{2}=-F\left(  \rho\right)  d\tau^{2}+G\left(  \rho\right)  d\rho
^{2}+\rho^{2}d\theta^{2}+\rho^{2}\sin^{2}\theta d\varphi^{2} \label{RNdual}%
\end{equation}
with%
\begin{align}
F\left(  \rho\right)   &  =-\frac{\left[  b^{2}M+b^{2}\sqrt{\left(
M^{2}-Q^{2}\right)  -Q^{2}\frac{\rho^{2}}{b^{2}}}\right]  ^{2}}{b^{2}\left(
b^{2}+\rho^{2}\right)  ^{2}},\\
G\left(  \rho\right)   &  =-\frac{\rho^{4}}{\beta^{2}Q^{2}\left[  b^{2}\left(
1-\frac{M^{2}}{Q^{2}}\right)  -\rho^{2}\right]  }.
\end{align}
Their geodesic equations are%
\begin{align}
\frac{d\varphi}{dr}  &  =\frac{1}{r^{2}}\frac{1}{\sqrt{\frac{1}{b^{2}}-\left(
1-\frac{2M}{r}+\frac{Q^{2}}{r^{2}}\right)  \left(  \frac{1}{a^{2}}+\frac
{1}{r^{2}}\right)  }},\\
\frac{d\phi}{d\rho}  &  =\frac{1}{\rho^{2}}\frac{1}{\sqrt{\frac{\beta^{2}%
}{B^{2}}\frac{b^{4}\left[  1-\frac{M^{2}}{Q^{2}}-\frac{\rho^{2}}{b^{2}%
}\right]  \left(  1+\frac{\rho^{2}}{b^{2}}\right)  ^{2}}{\rho^{4}\left[
\frac{M}{Q}+\sqrt{\frac{M^{2}}{Q^{2}}-1-\frac{\rho^{2}}{b^{2}}}\right]  ^{2}%
}-\frac{\beta^{2}Q^{2}b^{2}\left[  \frac{M^{2}}{Q^{2}}-1+\frac{\rho^{2}}%
{b^{2}}\right]  }{\rho^{4}}\left(  \frac{1}{A^{2}}+\frac{1}{\rho^{2}}\right)
}}.
\end{align}
It can be checked that Eqs. (\ref{RN}) and (\ref{RNdual}) satisfy the duality
relation (\ref{dual1})-(\ref{dual3}).

\section{Conclusions and outlook \label{Conclusion}}

In this paper, we introduce a duality between spacetimes. The duality relation
transforms both the metrics and the geodesics of two dual spacetimes to each
other. When two spacetimes are geodesically dual, their geodesics are related
by the duality transformation which also relates the metrics of the spacetimes.

The geodesic dual transformation transforms the geodesic equation to the
geodesic equation of the dual spacetime. Furthermore, the transformation that
transforms another kinds of dynamical equation to a dynamical equation on
another spacetime may lead to a new kind of duality of spacetimes. For
example, the geodesic is the trajectory of a free particle on spacetime
manifolds, while in quantum theory, various kinds of free particles are
described by various kinds of wave equations, scalar particles described by
scalar equations, spinor particles described by spinor equations, and so on.
This inspires us to seek dualities between spacetimes based on various kinds
of wave equations.

The geodesic corresponds\ only to free particles or free fields. It is worthy
to consider the duality relation based on various dynamical equations with
interactions. The geodesic is the solution of the geodesic equation.
Furthermore, various physical quantities are also solutions of various
dynamical equations. It is worthy to consider the duality relies on various
physical quantities. For example, for bound states, we can consider the
duality relation between bound-state eigenvalues. For scattering, the
information of the dynamics is embedded in the scattering phase shift
\cite{futterman1987scattering,li2018scalar,ahn2008black,rosa2012massive,glampedakis2001scattering,li2015heat,graham2009spectral,pang2012relation}%
, and we can consider the duality relies on phase shifts.

A further problem is the relation between various physical quantities on dual
spacetimes, e.g., the relation between eigenvalue spectra on dual manifolds.
In quantum field theory, the physical qualities we are interested in are all
spectral functions, such as the partition function (the global heat kernel)
and the effective action
\cite{vassilevich2003heat,dai2009number,dai2010approach}. The spectra and,
then, the spectral function is determined by the field equation on spacetimes,
so it is worthy to reveal the relation between the spectra and spectral
functions on dual spacetimes.

\acknowledgments

We are very indebted to Dr G. Zeitrauman for his encouragement. This work is supported in part by NSF of China under Grant
No. 11575125 and No. 11675119.






\end{CJK*}
\end{document}